\documentstyle[11pt,paspconf,epsf]{article}

\begin{document}

\title{The Hubble Deep Fields: North {\it vs}. South}

\author{Stephen D. J. Gwyn}
\affil{Department of Physics and Astronomy, University of Victoria,
    P.O. Box 3055, Victoria, BC, Canada V8W 3P6}

\begin{abstract}
Photometric redshifts have been calculated for 
the Hubble Deep Fields.
The redshift distributions of the fields differ; there is a large
excess of galaxies in the HDF-North in the redshift range
$0.4<z<1.2$.
The difference is consistent with the presence of a
weak cluster in the HDFN and is only slightly larger
than the cosmic variance in other fields of similar depths 
and with models of large scale structure.
\end{abstract}

% Keywords should be included, but they are not printed in the hardcopy.
\keywords{
% these are real PASP keywords 
galaxies: distances and redshifts, 
galaxies: photometry, 
large-scale structure of universe,
% In the specialised world of photometric redshifts,
% I'd also like my article to be indexed by the following:
photometric redshift techniques: template fitting,
empirical templates,
Hubble Deep Fields,
William Herschel Deep Field

}

\section{Introduction}

The Hubble Deep Field North (Williams {\it et al.}\ 1996)
resulted in dozens of papers on galaxy evolution.
All these papers treat the HDFN as a typical
field; the conclusions that are drawn are
assumed to hold for all fields.
With the advent of the HDF South (Williams {\it et al.}\ 1999)
it is possible to test this hypothesis.

\begin{figure}
\plotone{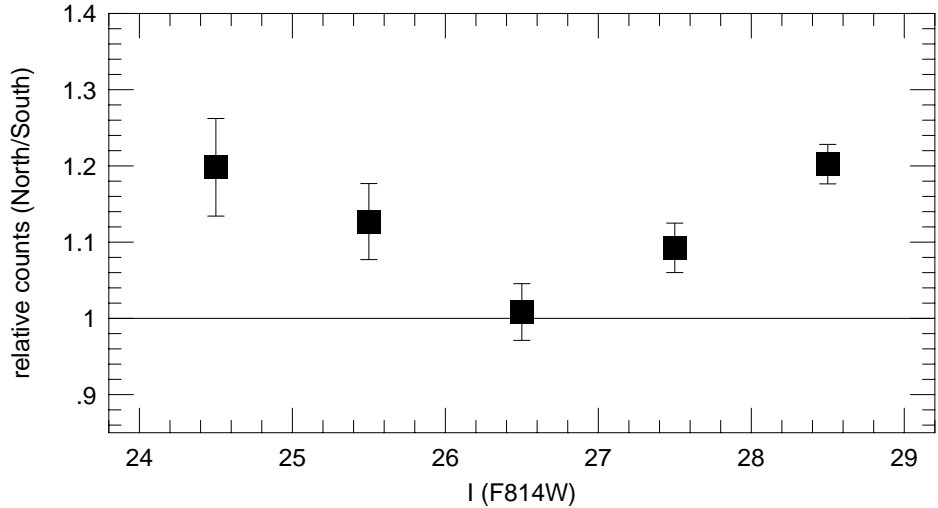}
\caption{Relative number counts for the Hubble Deep Fields. 
The ratio (North divided by South) of the F814W number counts
in each magnitude bin is shown. 
There are roughly 15\% more galaxies in the 
HDFN compared to the HDFS}
\label{fig:ncounts}
\end{figure}

The two fields do show some differences.
The number counts of 
Ferguson (1999, this volume) show
a that the HDFN holds 15\% more galaxies.
This excess is more visible when the differential
counts are plotted as shown in Figure \ref{fig:ncounts}.
Only a fraction of the galaxies in the HDFN 
and virtually none of the galaxies in the HDFS
have spectroscopic redshifts.
Thus the question ``where do these
excess galaxies in the HDFN lie?''
must be addressed with photometric redshifts.

\section{Catalogs and Photometry}
\label{sec:data}

The galaxy catalogs and photometry are generated 
using SExtractor (Bertin \& Arnouts, 1996) and additional
software written by the author.
First, SExtractor is run on the I band image of
the HDFN (version 2) and the HDFS (version 1) mosaics. 
SExtractor does an excellent job of deblending objects. 
The few errors it makes take the form of splitting 
single large, bright galaxies into fragments.
These are easily corrected by hand.

When determining photometric redshifts, 
one should measure the colours of galaxies through
the smallest feasible aperture.
Using a small aperture decreases the random error in the colours,
at the expense of introducing systematic shifts
if there is a colour gradient in the galaxy.
These systematic effects are actually desirable
since they generally take form a reddening
towards the centre.
Since reddening implies an increase in the amplitude of the
4000\AA~break, it is then easier to determine
a photometric redshift.
Using small apertures also minimises the chance
of contamination by other nearby galaxies.
However, it is desirable to construct a catalog using
a larger aperture to avoid the systematic errors.

For these reasons,
the final catalog contained galaxies with 
with $I_{ST}<$28\footnote{
All magnitudes in this article are given on the ST
magnitude system unless otherwise specified.
The ST magnitude system is defined
such that zero magnitude corresponds
to $F_\lambda=3.63\times10^{-12}~W\mu^{-1}$cm$^{-2}$
in all band passes. This is similar to the AB system
which is defined in terms of a constant value of $F_\nu$
but is convenient for comparing
magnitudes to template spectra which are usual given
in $F_\lambda(\lambda)$.
}
($I_{AB}<$27.2)
as measured through the 1.0 arcsecond aperture.
The colours measured through the smaller, 0.5 arcsecond, aperture
are used to determine photometric redshifts.
In both cases, pixels that lie within 
the isophotes of other nearby galaxies (as determined
using the segmentation image generated
by SExtractor) are excluded from the aperture.
Because the HDF frames in the different bands
are registered to within a fraction of a
pixel, the same pixels can be excluded on each frame.
This prevents colour contamination which
could affect the photometric redshifts.
Such contamination has particularly undesirable 
effects when faint U-band or 
B-band dropout galaxies lie near bright foreground galaxies.

\section{Photometric Redshifts}
\label{sec:phot}

The galaxies in the the Hubble Deep Fields 
span a large range in redshifts: 
The available spectroscopic redshifts in the Hubble Deep Fields,
although numerous below $z=1$ and in the range $2<z<3$, 
are spotty in the range $1<z<2$ and almost non-existent above $z=3$.
The various linear regression
photometric redshift techniques (e.g. Connolly {\it et al.}, this volume)
rely on a training set of spectroscopic redshifts.
These techniques, although effective at low redshifts 
where such a training set exists, are unreliable
where the spectroscopic coverage is sparse.
Therefore, the photometric redshifts in this article
are calculated using the template fitting technique.

The templates are constructed from the observed spectral
energy distributions of local galaxies.
The four spectra of Coleman, Wu \& Weedman (1980, CWW)
were used initially. 
It was found however, that many of the blue galaxies
in the Hubble Deep Fields are not well fit by even 
the bluest CWW spectrum.
This caused  moderate discrepancies when
the photometric redshifts were compared to the 
spectroscopic redshifts.
Therefore the CWW spectra
are supplemented with the SB3 and SB2 spectra
from Kinney {\it et al.}\ (1996) to form
the basis of the template set.
From this basis set of six spectra,
intermediate templates are constructed by
interpolation for a total of 51 templates.

These templates are redshifted at intervals
of 0.02 in $\log_{10}(z)$. 
Spacing the templates in $\log z$ is an 
improvement over the more usual linear spacing.
It allows,
for the same total number of templates,
tighter coverage at low redshift (where
it is most needed) at the small sacrifice of
sparse coverage at high redshift (where
it is not needed).
The spectra are corrected for
intergalactic absorption as prescribed by Madau (1995)
After redshifting, the templates are multiplied
by the response function of the UBRI filters
to produce fluxes at the central wavelength of each filter.
These fluxes are converted to magnitudes to form
the final templates.

Each template is compared to the 
observed galaxy magnitudes in turn and a
$\chi^2$ is determined:
\begin{equation}
\chi^2=\sum_{i=1}^{N_{filters}} {(M_i-T_i-\alpha)^2 \over \sigma_{M_i}^2},
\label{eq:chi2}
\end{equation}
Where 
$M_i$ is the observed magnitude of the galaxy,
$\sigma_{M_i}$ is the magnitude uncertainty,
$T_i$ is the template magnitude, and
$\alpha$ is a normalisation factor
that corrects the templates
to the apparent magnitude of the galaxy.
The optimal normalisation factor is determined by minimising
equation (1) with respect to $\alpha$.

In many cases the galaxy is undetected in one or more 
of the band passes.
This can occur when the galaxy when a galaxy is
at high redshift and its UV flux has been absorbed
by the IGM (the U-band dropouts).
However, this can also occur with intrinsically
faint, low redshift galaxies. 
The situation is handled by replacing the relevant term of the sum
in equation (1).
If the magnitude predicted by template is less than
the magnitude limit in that band, the term is replaced with zero.
If this isn't case, on the other hand, 
the term is replaced with 
\begin{equation}
{ M_{\rm limit}-T_i-\alpha \over \sigma_{T_i}^2 }
\end{equation}
where
$M_{limit}$ is the magnitude limit of the image in question.
The weighting factor, $\sigma_{T_i}$, 
is the uncertainty that the galaxy's magnitude
would have if it was visible in that bandpass.

\begin{figure}
\plotone{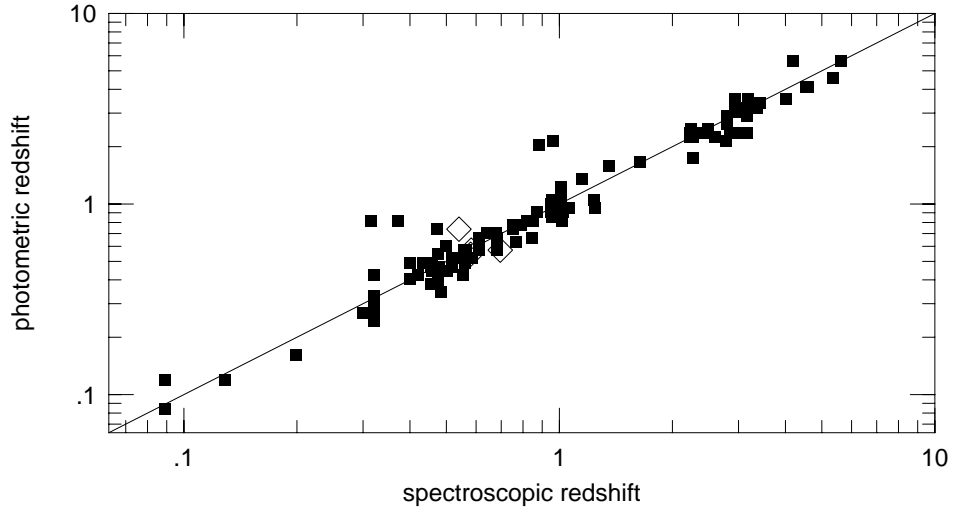}
\caption{A comparison of photometric and spectroscopic redshifts.
The filled squares represent galaxies in the HDF North; those
in the HDF South are shown by open diamonds.
 }
\label{fig:zz}
\end{figure}

As a check on the accuracy of the technique,
photometric redshifts were calculated by the above
method for galaxies in both Hubble Deep Fields
which have spectroscopic redshifts.
The comparison is shown in Figure \ref{fig:zz}.
The redshift uncertainties scale with z.
The typical relative error in the photometric redshifts
is $\sigma_z/z=11\%$. 

\section{The Comparison}
\label{sec:comp}
\begin{figure}
\plotone{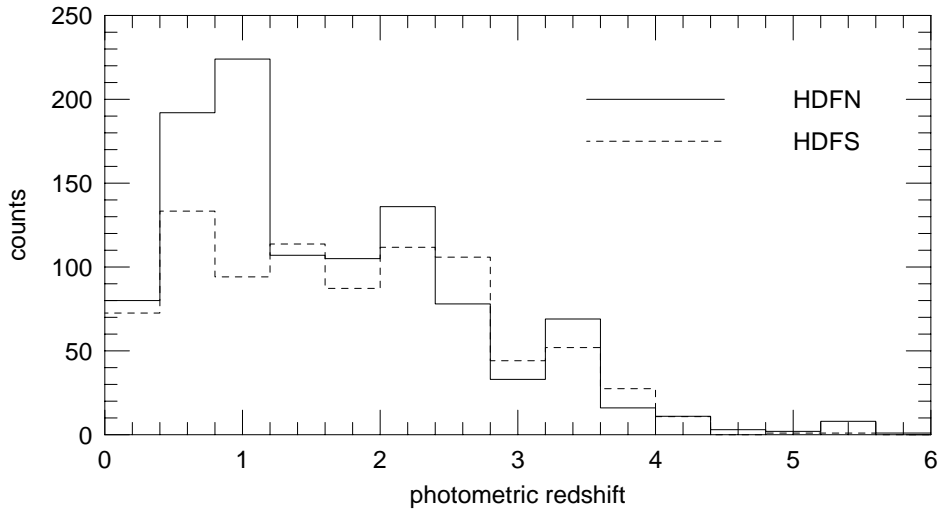}
\caption{Photometric redshift distributions for the HDF North (solid line)
and the HDF South (dashed line).
 }
\label{fig:zhist}
\end{figure}

The photometric redshift technique described in section
3 was applied to the photometric catalogs
described in section 2.
The resulting redshift distributions for the HDF North
and South are shown in Figure \ref{fig:zhist}.
The two redshift distributions are not the same.
The Kolmogorov-Smirnov test gives the probability
of the two distributions being the same as $1.2 \times 10^{-6}$.
The redshift distributions are most different
in the redshift range $0.4<z<1.2$.

\begin{figure}
\plotone{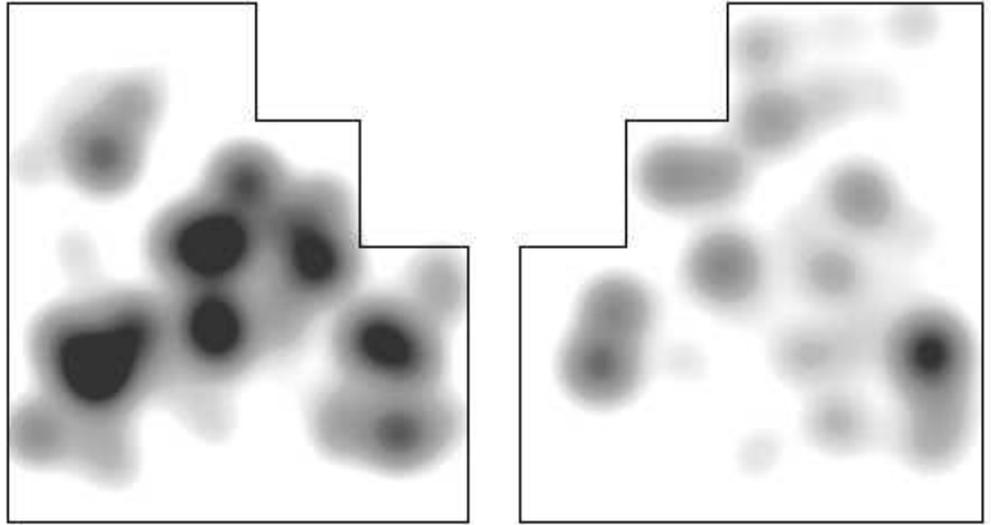}
\caption{I band light in the HDFN (left) and HDFS (right) coming from galaxies
with photometric redshifts in the range $0.4<z<0.8$.
The image has been convolved with a 6 arcsecond radius Gaussian.
Note the concentration of light near the centre of the North image
which is not present in the South}
\label{fig:blob}
\end{figure}

It is tempting to ascribe the differences
in the Hubble Deep Fields to a structure
present in the North but not in the South.
Indeed, there is a pronounced spike in the
spectroscopic redshift distribution of
the HDFN at $z=0.475$ (Cohen {\it et al.}\ 1996).
Figure \ref{fig:blob} shows the I band images
of the Hubble Deep Fields. Only light from galaxies
with photometric redshifts in the range $0.4<z<0.8$
is shown; the other galaxies been masked out.
%(see {\tt http://astrowww.phys.uvic.ca/grads/gwyn/pz/dice.html})
The images have been convolved with 
Gaussian profile ($\sigma=6$ arcseconds).
The left image shows a large concentration of light
in the HDFN that is not present in the South.

The two-point angular correlation function was
computed for various redshift slices.
Note that it is impracticable to calculate the
angular correlation function for slices much narrower
than about $\Delta z=.4$ without running
into problems with small number statistics.
Also, because of the uncertainties on the
redshifts, it would be difficult to compute
a reliable spatial (as opposed to angular) correlation function.
The correlation functions for both fields
were compared for each slice. For most redshift
slices, they showed no difference within the errors.
The only exception was the in the $0.4<z<0.8$ redshift
slice, where galaxies in the HDFN were
significantly more clustered than in the HDFS.
The spatial scale of the structure (the HDF is 
$\sim$ 1 Mpc across at that redshift) and the number of galaxies involved
($\sim 50$ more galaxies in the North
than in the South) suggest a very poor cluster
or a very rich group.

More generally, the differences in the redshift distributions
could be due to cosmic variance in the large
scale galaxy distribution. 
This hypothesis was tested empirically in the following manner:
The  William Herschel Deep Field (WHDF, Metcalfe {\it et al.}\ 1999 
in press) extends to $B=28$ and has good coverage
in the UBRIHK bands. It covers roughly 40 square arcminutes.
The WHDF was divided into 9 separate areas, each the 
same size as the Hubble Deep Fields. 
%The analysis was not complete at the time of this writing 
%and only preliminary results are presented here. 
The field to field 
variance was found to be 10\% (rms), smaller than, but not 
inconsistent with, the difference between the HDFN and
HDFS.

N-body simulations computed by Stadel (private
communication) indicate the variance in the mass
distribution along lines of sight comparable the HDF
are about 20\% out to $z=1$. Assuming that galaxies are linearly
biased (Kauffman, 1998), this should
translate into a similar variance in the redshift
distributions in the Hubble Deep Fields. Again
this is slightly smaller than, not but not 
inconsistent with, the difference between the two
redshift distributions below $z=1$ as seen in Figure \ref{fig:zhist}.

%\acknowledgements

%Thanks is due to my supervisor, F. D. A. Hartwick, for his continuing 
%support. 

\end{document}